Vaschenko V.M.*, Loza Ye.A.**

# Influence of the stratosphere temperature on ozonosphere optical characteristics and instrumental problems of total ozone content remote measurements

*In this paper we investigate stratosphere temperature impact on remote ozone satellite and ground-based optical observations. High correlation between stratospheric temperature and instrumentally determined total ozone content requires taking into account temperature dependency of ozone absorption and scattering indexes and of other atmosphere characteristics for inverse ozone observations problem solution. The assumption that the majority of observed ozone anomalies and trends are caused by atmosphere temperature change is made.*

*State Ecological Academy, Kyiv, Ukraine, E-mail: Daniilko@mail.ru
**State Ecological Academy, Kyiv, Ukraine, E-mail: Loza@bmyr.kiev.ua

## Introduction

Remote passive investigations of the Earth atmosphere state and content are based on spectrometric analysis of scattered in the atmosphere solar radiation. Its parameters depend on spectral scattering coefficients of atmospheric components - oxygen, nitrogen, small atmospheric components including ozone and aerosol.

Optical radiation scattering characteristics by atmospheric components depend on many factors, e.g. their concentration, external pressure, atmosphere temperature [1], external electric and magnetic fields [2], solar constant variations [3], etc. Moreover there are effects connected to fractal fluctuations in the atmosphere [4] and overlapping of absorption, radiation and scattering spectra of the atmospheric components.

Remote atmosphere spectrometry considers these and other phenomenon by atmospheric models created for regional scale with some degree of account for seasonal variations. The accuracy of experimental results obtained by such models application is satisfactory for latitudes up to 70º due to a specific quantity of contact (chemical) experimental data on small atmospheric components concentration. However, the accuracy of ozone observations may reduce due to anomalous atmospheric phenomenon or insufficient atmosphere experimental investigations in high latitudes.

Therefore, taking into account ecology, climate and socio-political importance of the Earth ozone layer, this paper considers interconnection between temperature variations and anomalies and trends of instrumentally determined total ozone content (ITOC). The conclusions and suggestions for other probable phenomenon are made.

**Problem formulation**

It is known that molecular absorption spectral band form, in particular its full width at half maximum and maximum height, depend on external conditions - and first of all on temperature. The spectral band broadening may be inhomogeneous (first of all caused by Doppler effects which strongly depends on gas molecules velocity distribution, i.e. on temperature) and homogeneous (mainly due to impact broadening that depends on temperature and pressure). These two broadening types are always observed together in nature and the spectral band form is described by Voigt profile [5, 6]. Spectral bands of molecular absorption may have very complex form due to overlapping of different oscillatory and rotational energy levels and produce a sophisticated optics and thermodynamics problem.

Moreover, apart from classical effects we should also consider non-linear interaction cross-section increase with temperature increase and absorption cross-section temperature hysteresis phenomenon, first of all due to hysteresis changes in chemical-and-physical atmosphere content due to temperature change [7].

Ozone observations are also influenced by temperature change of other atmosphere components optical properties due to the same temperature effects in other molecular and atomic gases and ions. Moreover temperature changes in atmosphere aerosol are much harder to predict, e. g. due to possible phase changes [7] and chemical compound change due to temperature-dependent reactions with external environment.

The natural temperature range in the Earth atmosphere is greater than 100 degrees - from +50 ºC at the equator to -70 ºC at the poles. This leads to significant ozone scattering coefficients change and therefore to non-linear increase in ITOC bias errors. The registered temperature records for the last century are -89.2 ºC ("Vostok" station, Antarctic, 1983) and +57.8 ºC (Libyan desert, Libya, 1922). At height of ozone maximum (20-30 km ) the stratospheric temperature is less than near-surface by 45-75 degrees and the pressure drops from 1000 to 20-100 mBar.

**Ozone absorption and scattering coefficients temperature change**

Many authors stress the importance of knowing the experimental temperature dependency for ozone scattering and absorption effective cross-sections for accurate determination of atmosphere transparency in visible and UV spectrum ranges [8-11].

Changes in ozone absorption coefficients in visible range at temperature change by 80 degrees are estimated up to 40% at edges of the absorption band [12] and up to 10% at its maximum [13]. Considering the fact that absorption coefficients used in atmospheric models differ from the last experimental results [14] there is an high demand for detailed laboratory investigations of ozone optical properties dependency on temperature and pressure and also for synchronous atmosphere ozone and temperature observations and for effective theoretical algorithm of these data interpretation.

Theoretical calculations [1] show that atmosphere transparency and, respectively, absorption and scattering coefficients are directly-proportional to both temperature and pressure at near-ground atmosphere layer. Especially these changes are essential in UV range, where ITOC determination error may reach 15% at temperature change by 40 degrees and about 6% at pressure change by 40 mBar [1].

For latitudes above 70º due to climate conditions influence (and first of all the stratosphere temperature) an unaccounted earlier bias error appears that according to ground-based [15] and satellite [16] experimental investigations may be estimated 10% at minimum. It may be avoided only by parallel atmosphere temperature measurements. It is well-known that Dobson spectrometer and others standard devices for ozone concentration measurements give significantly underestimated values of ITOC in case atmosphere temperature reduction at the height of ozone maximum [17].

Using this dependency a method based on ozone absorption spectral band wings intensity measurements synchronous to total ozone content measurements by spectral band maximum was proposed [18-20].

## Temporal anomalies and trends
## of instrumentally determined total ozone content

Such ITOC temperature dependency leads to correlation of long-term ITOC reduction trend [21] with temperature reduction on decades scale [22-24]. For the height of ozone maximum a high positive correlation of ITOC and temperature is observed for quasi-two years and half-year ITOC oscillations according to spectrometry [25-27] and LIDAR data [28]. Short-period correlation of ITOC and temperature is also observed at 13-27 days scale [29-31]. Similar phenomenon were also observed for other gases [32].

At heights 30-80 km anti-correlation of ITOC and temperature is found, while for other heights - a high positive correlation with no time delay is observed [25-27, 33-36]. ITOC and temperature dynamics is described by equal fractal dependencies [4].

For some experiments the ITOC and near-ground temperature correlation reaches 0.9 [37, 38]. However such dependency may not be considered reliable because near-ground temperature is not linearly connected to the stratospheric temperature.

Therefore a large quantity of investigations show that at all timescales from days to decades a high correlation between ITOC and stratosphere temperature is observed pointing at their close interconnection. Here ozone integral spectral scattering and absorption coefficients variation with temperature change may be misinterpreted as total ozone content change. This is also true for remote optical investigations of other small atmospheric components.

## Spatial anomalies of instrumentally determined total ozone content

The results of 14-years atmosphere observation by TOMS (Total Ozone Mapping Spectrometer) and MSU (Microwave Sounding Unit) at 150-50 mBar height showed stable high temporal-spatial correlation between stratospheric temperature and ITOC both for local phenomenon and for global trends [39, 40]. The same dependency was found for sudden stratosphere heating in Arctic in 2002, revealed by space Fourier-spectrometer MIPAS (Michelson Interferometer for Passive Atmospheric Sounding) [41]. The same effect was also revealed by other ground-based and satellite optical measurements [42].

Moreover, in different regions the value of the correlation is constant. It changes depending on geographic location of the observation site [43].

From 30 November to 1 December 1999 a "mini-ozone hole" was observed over Europe with its maximum coinciding with temperature minimum in the tropopause [44]. The same localized manifestation of ITOC and temperature coupling was observed in October 1987 and November 1999 in polar regions [45].

Investigation of ozone anomalies over Europe during winter 1991-1992 revealed correlation between temperature and instrumentally determined ozone partial pressure in the atmosphere [46]. The same phenomenon was observed for the edge of the Polar stratospheric vortex and the South oscillation [47]. Synchronous observation at McMurdo station at Antarctic Peninsula also show high positive correlation between ITOC and temperature [48].

Localization of low stratosphere temperature over Antarctic is explained in [49]. Therefore we may make a conclusion that the reduction in ozone layer absorption ability in polar regions during polar winter may be caused not by real atmospheric ozone quantity change, but by stratosphere temperature influence on ozone molecules and other atmospheric components scattering and absorption coefficients.

The phenomenon of ITOC dependency on temperature and other atmospheric components characteristics will manifest for all the optical methods based on measurement of optical radiation absorption or scattering by ozone. Due to temperature change all the methods including UV-spectrometry and LIDAR investigations will correlate with one another but despite this will have a high uncontrolled bias error.

**Time-delayed temperature phenomenon in ozone layer**

We should also stress that ITOC and temperature correlation is non-linear [50]. Moreover, there are inert temperature phenomenon.

The ozone layer state significantly influences the atmosphere temperature - the more ozone absorption coefficients and ozone content are the more energy it absorbs changing the atmosphere temperature. Moreover the time-delayed phenomenon are connected to temperature dependency of ozone chemical reactions [22]. Also ozone isotope content changes depending on stratosphere temperature [51, 52].

On short time intervals under solar UV-radiation flux change at low latitudes according to the data of SBUV (Solar Backscattered Ultraviolet Instrument) and SAMS (Stratospheric and Mesospheric Sounder) installed at Nimbus 7 satellite a correlation of ITOC and solar flux variation was found to be 0.3 to 0.6 with phase shifts from 3 to 13 days at different heights [53-55]. Also 27-days variations in spatial distribution of ITOC are found to be connected to Sol rotation [56].

**Atmosphere dynamics and atmosphere aerosol impact**

During sand storms in deserts it was found that ITOC correlates with atmosphere aerosol state [57]. Based on in-year synchronous ITOC and temperature variations investigation their connection to atmosphere aerosol state was found [58, 59]. There is an suggestion that temperature and aerosol state change together with solar activity variations played the most important role in reduction of ITOC in 1979-1993 [60, 61].

A reliable interconnection of ITOC and wind intensity was found [62, 63], which was associated by some authors with parallel stratosphere cooling [64]. Atmosphere dynamics influence on ITOC was also observed [65-69].

## Conclusions

1. The temperature of the lower stratosphere significantly influences instrumentally determined total ozone content due to ozone absorption and scattering coefficients temperature dependency. Contemporary ozone measurement methods do not consider temperature impact during total ozone content determination leading to bias errors at least 15-20%, especially at polar regions. Therefore careful experimental investigations of ozone scattering and absorption coefficients temperature dependency determination is required to include it in theoretic models.

2. High positive correlation between instrumentally determined total ozone content and stratosphere temperature is observed both in time and in space. The correlation coefficients of instrumentally determined total ozone content and stratosphere temperature are found to be 0.6 to 0.9 in different papers. This means that bias error of instrumentally determined total ozone content may be over 90%.

3. Anomalous or seasonal stratosphere temperature reduction leads to phenomenon of instrumentally determined total ozone content reduction. As a result, the phenomenon of "ozone holes" may be explained by optical-and-temperature phenomenon due to change of ozone absorption and scattering coefficients and not ozone molecules quantity change. This is especially important for Antarctic and Arctic where Sol zenith angles are large and stratosphere temperature is very low in winter.

4. Measured by remote methods instrumental values of total ozone content may also change due to atmosphere dynamics and temperature or seasonal change in properties of other atmospheric components, first of all - aerosol.

5. In order to avoid bias error and to understand real physical-and-chemical nature of ozone holes and planetary waves measurement complexes for synchronous investigation of total ozone content, aerosol characteristics and temperature must be developed. Also theoretical apparatus of ozone models should be improved to accommodate these phenomenon.